\documentclass[fleqn,usenatbib]{mnras}

\usepackage[T1]{fontenc}

\DeclareRobustCommand{\VAN}[3]{#2}
\let\VANthebibliography\thebibliography
\def\thebibliography{\DeclareRobustCommand{\VAN}[3]{##3}\VANthebibliography}

\usepackage{graphicx}	
\usepackage{subcaption}
\usepackage{latexsym,amsmath,amssymb,lmodern,url}
\usepackage{natbib}
\usepackage{courier}
\hypersetup{colorlinks=true}
\usepackage{bm}
\usepackage{dcolumn}
\usepackage{color}
\usepackage{xcolor}
\usepackage{comment}
\usepackage{soul}

\def\vlos{v_{\text{los}}}
\def\vrad{v_{\text{rad}}}
\def\vtan{v_{\text{tan}}}

\title[Sgr Kinematics: 5D Analysis, 6D Methodology]{Kinematics of the Sagittarius Dwarf Spheroidal core: A 5D Analysis for a 6D Methodology with Gaia DR3}

\author[I. S. Goldstein et al.]{
Isabelle S. Goldstein,$^{1}$\thanks{E-mail: isgoldstein@tamu.edu}
Louis E. Strigari$^{1}$
\\
$^{1}$Mitchell Institute for Fundamental Physics and Astronomy, Department of Physics and Astronomy, Texas A \& M University, College Station, TX 77843, USA
}

\date{Accepted XXX. Received YYY; in original form ZZZ}

\pubyear{\the\year{}}

\begin{document}
\label{firstpage}
\pagerange{\pageref{firstpage}--\pageref{lastpage}}
\maketitle

\begin{abstract}
Using Gaia DR3 data, we examine the kinematics of the central core of the Sagittarius (Sgr) dwarf spheroidal galaxy using data which includes proper motions and line-of-sight velocities for member stars in addition to their projected positions. We extract a sample of bright stars that are high-probability members of Sgr. We model the distances to these stars, which is the only missing phase-space component measurement from our 5D sample, highlighting how their corresponding uncertainties propagate to affect the kinematics. Using line-of-sight velocity data only, which are not affected by the distance uncertainties, and assuming a jeans-based equilibrium analysis, we obtain a velocity anisotropy of $\beta_a = -2.24 \pm 1.99$, which implies a system with tangentially-biased orbits. With the full 5D data, we project that the data will significantly improve upon measurements of the log-slope of the dark matter density profile and the stellar velocity anisotropy. 
Tests with mock distance data show an improvement of anisotropy errors of approximately an order of magnitude, and log slope at the half light radius of approximately half an order of magnitude.
\end{abstract}

\begin{keywords}
Galaxy: kinematics and dynamics – Cosmology: dark matter – stars: kinematics and dynamics
\end{keywords}




\section{Introduction}\label{sec:intro}

\par The Sagittarius (Sgr) dwarf spheroidal (dSph) is one of the most luminous satellite galaxies orbiting the Milky Way (MW)~\citep{Ibata1994Natur.370..194I}. It is unique from other dSphs in that it has both a central core and long, extended tails that indicate tidal interaction with the MW~\citep{2010ApJ...714..229L}. The photometric and kinematic properties of the central core and tidal tails, combined with simulations of its orbit, provide insight into the nature of the Sgr progenitor before it was accreted into the MW, and provide estimates for its dark-to-luminous mass ratio~\citep{2010MNRAS.408L..26P,Lokas2010ApJ...725.1516L}. Sgr-like progenitors have also been identified in large-volume cosmological simulations~\citep{2024A&A...687A..82L}, which provide insight into its evolution within a cosmological context. 

\par New astrometric data from Sgr, in particular in the form of proper motions from the Gaia satellite, have provided a wealth of information on its properties. The proper motion of Sgr constrains the nature of its orbit within the MW~\citep{2018A&A...616A..12G}. Stars associated with the stream may be isolated in phase space, and can be used to constrain its chemical abundance and the nature of its orbit~\citep{2020A&A...635L...3A,2020ApJ...889...63H,2020MNRAS.497.4162V,2021MNRAS.508L..26P,2022ApJ...940L...3W,2023ApJ...946...66L,2024ApJ...963...95C}. Within the central core, astrometric data provide new insights into the internal structure of M54, the globular cluster at the center of Sgr, and the structure of the main body of Sgr~\citep{2020MNRAS.495.4124F,2024MNRAS.532.3713A}.  

\par Combined with the astrometric data from Gaia, line-of-sight velocities for Sgr stars ~\citep{Ibata1997AJ....113..634I,Frinchaboy2012ApJ...756...74F,2012MNRAS.427.2647M} provide the full 3D velocity space structure, and an estimate of the internal kinematic properties of the core, such as its rotation and velocity dispersion.~\citet{2020MNRAS.497.4162V} combine radial velocity data sets with Gaia DR2 astrometric data to estimate the 3D structure, and use this to constrain the orbit and the equilibrium nature of the core.~\citet{2021ApJ...908..244D} use machine learning methods to obtain a sample of Sgr stars with radial velocities and distances, and estimate the rotation and internal structure of the core.  Though the aforementioned studies are motivated by examining the 3D velocity structure of the Sgr core, given the data sets used, they ultimately relied on modeling methods to obtain the line-of-sight velocities for stars with astrometric data. 

\par With new high-quality data sets, it is timely to ask if they can be used to better constrain the dark matter in Sgr. The mass distributions of dSphs have long been understood to be excellent probes of the properties and nature of dark matter~\citep{1983ApJ...266L..11A}. Traditionally, high-quality measurements of the photometry and the stellar line-of-sight velocities provide estimates of the integrated dark matter mass of the dSphs. However, a measurement of the slope of the central density profiles, which provides an important test of Navarro-Frenk-White (NFW) profiles predicted by Cold Dark Matter (CDM) theory, has remained elusive with these data sets. Some analyses find constant central density cores~ \citep{Gilmore2007ApJ...663..948G, Walker2011ApJ...742...20W, Agnello2012ApJ...754L..39A,Amorisco2012MNRAS.419..184A, Amorisco2012ApJ...756L...2A, Amorisco2013MNRAS.429L..89A} while others find compatibility with NFW profiles~\citep{Strigari2010MNRAS.408.2364S,Breddels2013MNRAS.433.3173B, Jardel2013ApJ...775L..30J,Strigari2017ApJ...838..123S}. 

\par Additional phase space measurements beyond line-of-sight velocities has long been discussed as a possible method to better constrain dSph dark matter density profiles. The use of proper motions have been discussed by several authors~\citep{Wilkinson2002MNRAS.330..778W, Strigari2007ApJ...657L...1S, Read2017MNRAS.471.4541R,Guerra:2021ppq}, as well as the use of distances to individual stars~\citep{2014MNRAS.440.1680R}. The first measurement of the tangential velocity dispersion in a dSph was made for Sculptor by combining archival {\it HST} imaging with {\it Gaia} DR1 imaging~\citep{Massari2018NatAs...2..156M}. However the uncertainties on the velocity dispersions preclude the ability to distinguish between cored and cusped dark matter density profiles~\citep{Strigari2018ApJ...860...56S}. More recently, also by combining archival {\it HST} imaging with {\it Gaia} DR2 imaging, the first internal stellar tangential velocity dispersion measurement has been made for Draco~\citep{Massari2019arXiv190404037M,2024ApJ...970....1V}.

\par Though the dark matter distributions of many dSphs have been examined in detail, there have been relatively few studies of Sgr, particularly in the context of equilibrium models~\citep{Lokas2010ApJ...725.1516L}. This is mainly because it is not clear that an equilibrium model is appropriate, even for the central core of Sgr. For example~\citet{2022ApJ...941..108W} undertook a detailed study addressing this question using axisymmetric Jeans models, and showed that the core of Sgr may be reasonably well-described as in equilibrium. On the other hand, using DR2 data,~\citet{2020MNRAS.497.4162V} find that the core is likely not in equilibrium, and therefore as an extension likely not well-described by Jeans-based dynamical models. 

\par Determining whether an equilibrium model works for Sgr also has important implications for indirect dark matter detection. Indeed, there has been recent identification of a gamma-ray source in the direction of Sgr~\citep{2020ApJS..247...33A}, which given uncertainties, may be compatible with a dark matter annihilation model~\citep{2023MNRAS.524.4574E,2024JCAP...10..019V}. Estimates of the dark matter mass from stellar kinematics is important for efforts to constrain, or possibly detect, the annihilation cross section scale. 

\par With the above motivation, in this paper we obtain a sample of Sgr member stars with full 3D velocity information. We use the Gaia DR3 data set, which in addition to astrometric data provides line-of-sight velocities for stars with magnitude $G \lesssim 16$. We model the distances to stars, and use this to determine what information may be extracted on the velocity anisotropy and the dark matter density profile. We model the three components of the velocity dispersion with spherical Jeans-based models for cored and cusped dark matter profiles, and discuss the prospects for distinguishing between them with future data sets. We examine the impact of uncertainties for the distances of stars, and discuss how future measurements of such distances will improve upon the constraints we consider.  

\par This paper is organized as follows: \S\ref{sec:analysis} describes the coordinate system (\S\ref{subsec:coordTrans}), Jeans analysis (\S\ref{subsec:jeansAnalysis}), and setup for the Bayesian analysis (\S\ref{subsec:bayesAnalysis}). \S\ref{sec:data} describes the data selection and mock distances. \S\ref{sec:results} discusses the Jeans analysis results and their interpretation for use on data with real distances, and \S\ref{sec:conclusion} is concluding thoughts. 


\section{Analysis}\label{sec:analysis}
In this section, we discuss the coordinate transformations used in our analysis, followed by an overview of Jeans modeling, and then a discussion of the Bayesian statistical analysis. 

\subsection{Coordinate transformation}\label{subsec:coordTrans}
Gaia provides precision measurements of the positions of stars in terms of their right ascension (RA) and declination (DEC) $(\alpha, \delta)$, and the proper motions along these directions $(\mu_{\alpha*}, \mu_\delta)$ where $\mu_{\alpha*}=\mu_\alpha \cos\delta$. Line-of-sight (LOS) velocities are also measured for the brightest sample of stars, with magnitudes $G \lesssim 16$. Let $(\alpha_0, \delta_0, D_0)$ be the RA, DEC, and distance to the center of Sgr, which we assume to correspond to the center of mass of the galaxy. We can define the radius vector in the plane of the sky, $\mathbf{R}$, as the 2D vector from the center of Sgr to a given star. 
We will consider stellar velocities broken into three components: 1) directed along the LOS, 2) directed parallel to $\mathbf{R}$, which we denote as {\it rad}, and 3) directed tangential to $\mathbf{R}$, which we denote as {\it tan}. Positive ${\bf R}$ points radially outward and along the LOS points away from the observer; the positive direction in {\it tan} is then defined with the right-hand rule. 

Given these definitions, we can follow the coordinate transformation given in \citet{2002AJ....124.2639V,2021ApJ...908..244D}. Referring to the coordinate definitions from~\citet{2002AJ....124.2639V} and~\citet{2001AJ....122.1807V}, we can define the angular position of a star in the plane of the sky by 
\begin{equation}\label{eqn:angDistDefns}
\begin{split}
        \rho &= \arccos\left( \cos\delta \cos\delta_0 \cos(\alpha-\alpha_0) + \sin\delta \sin\delta_0\right) \\
        \phi_\ast &= \arctan2\left( 
        \frac{\sin\delta \cos\delta_0 - \cos\delta \sin\delta_0 \cos(\alpha - \alpha_0)}{-\cos\delta \sin(\alpha-\alpha_0)}
        \right)
\end{split}
\end{equation}
where $\rho$ is the angular separation from the center of Sgr and $\phi_\ast$ is the position angle measured counterclockwise with respect to the axis running in the direction of decreasing right ascension at constant declination. Then the radial and tangential velocity components in the plane of the sky, in units km/s, can be written as, 
\begin{equation}\label{eqn:velcomps}
\begin{split}
    v_\text{rad} &= \left(\frac{4.7403885}{\text{mas yr}^{-1}}\right) \left(\frac{D}{\text{kpc}}\right) [\mu_{\alpha*} \sin\Gamma + \mu_\delta \cos\Gamma] \\ 
    v_\text{tan} &= \left(\frac{4.7403885}{\text{mas yr}^{-1}}\right) \left(\frac{D}{\text{kpc}}\right) [\mu_{\alpha*} \cos\Gamma - \mu_\delta \sin\Gamma]
\end{split}
\end{equation}
where 
\begin{equation}\label{eqn:Gamma}
\begin{split}
    \cos \Gamma &=  [\sin \delta \cos \delta_0 \cos (\alpha - \alpha_0) -\cos{\delta}\sin{\delta_0}]/  \sin \rho \\ 
    \sin \Gamma &= [\cos \delta_0 \sin (\alpha - \alpha_0)] / \sin \rho. 
\end{split}
\end{equation}
Note that $v_{los}, v_{rad},$ and $v_{tan}$ here are respectively the $v_1, v_2,$ and $v_3$ of \citet{2002AJ....124.2639V}. 

We will be interested in the intrinsic internal mean velocity and velocity dispersion for our sample of stars in Sgr. Therefore we must correct the measured velocities for the center of mass (COM) motion. The contribution of the COM motion to the velocity of a star is given by \citep{2021ApJ...908..244D}
\begin{equation}\label{eqn:COMmotion}
\begin{split}
    \vlos &= v_0\sin\rho \cos(\phi_\ast-\theta_t) + v_{\text{LOS},0}\cos\rho \\
    \vrad &= v_0\cos\rho\cos(\phi_\ast-\theta_t) - v_{\text{LOS},0}\sin\rho \\
    \vtan &= -v_0 \sin(\phi_\ast - \theta_t),
\end{split}
\end{equation}
where $v_{\text{LOS},0}$ corresponds to the overall systemic motion of Sgr along the LOS. The COM velocity projected on the plane of the sky is given by magnitude $v_0$ and angle $\theta_0$, defined in the same way as $\phi_\ast$: 
\begin{equation}
    \begin{split}
        v_0 &= \left(\frac{4.7403885}{\text{mas yr}^{-1}}\right) \left(\frac{D_0}{\text{kpc}}\right) \sqrt{\mu_{\alpha*0}^2 + \mu_{\delta0}^2} \\
        \theta_0 &= \arctan2\left(\frac{\mu_{\delta0}}{-\mu_{\alpha*0}}\right) .
    \end{split}
\end{equation}

\subsection{Jeans Analysis}\label{subsec:jeansAnalysis}

We utilize the spherical Jeans equation to connect the observed velocity dispersions to the underlying gravitational potential, ${\bf \Phi}$. We define $r$ as the 3-dimensional radial coordinate, and $R$ as the 2-dimensional projected radial coordinate. Note that $R$ is the magnitude of the vector ${\bf R}$ introduced above. In spherical coordinates and under the assumption of a time-independent spherical system the Jeans equation can be written as follows~\citep{GalacticDynamics}
\begin{eqnarray}
\label{eqn:jeans}
\frac{d{\bf\Phi}}{dr} &=& - \frac{G M(r)}{r^2} \nonumber \\
&=& - \frac{1}{\nu(r) } \frac{d}{dr}  \left[ \nu(r) \overline{u_r^2}(r) \right] - 2 \frac{\beta_a (r)  \overline{u_r^2}(r)}{r}.
\end{eqnarray}
where $\nu(r)$ is the stellar density profile, and the components of the intrinsic velocity dispersion are 
$\overline{u^2_r}(r), \overline{u^2_\theta}(r), \overline{u^2_\phi}(r)$. Note that the assumptions of a time-independent and spherical system mean we are not modeling streaming or tidal disruption. This also disregards the observed mean ellipticity $\epsilon\sim0.64$ measured for Sgr~\citep{McConnachie_2012}. 
The stellar velocity anisotropy, $\beta_a$, is 
\begin{equation}\label{eqn:betaA}
\beta_a(r) \equiv  1 - \frac{ \overline{u_\theta^2}(r) }{\overline{u_r^2}(r)}. 
\end{equation}
With the assumption of spherical symmetry we have $\overline{u_\theta^2}(r) = \overline{u_\phi^2}(r)$. The velocity dispersions in the {\it los}, {\it rad}, and {\it tan} directions are then given by~\citep{2007ApJ...657L...1S}

\begin{equation}\label{eqn:veldisplos}
    \sigma^2_{los} = \frac{2}{\Sigma(R)} \int_R^\infty \left(1-\beta_a(r)\frac{R^2}{r^2} \right) \frac{\nu(r)\,\overline{u_r^2}(r)\, r}{\sqrt{r^2-R^2}}dr,
\end{equation}

\begin{align}\label{eqn:veldisprad}
    \sigma^2_{rad} = \frac{2}{\Sigma(R)} \int_R^\infty 
    &\left(1 -\beta_a(r)+\beta_a(r)\frac{R^2}{r^2} \right) \nonumber \\
    &\times \frac{\nu(r)\,\overline{u_r^2}(r)\, r}{\sqrt{r^2-R^2}}dr,
\end{align}

\begin{equation}\label{eqn:veldisptan}
    \sigma^2_{tan} = \frac{2}{\Sigma(R)} \int_R^\infty \left(1-\beta_a(r)\right) \frac{\nu(r)\,\overline{u_r^2}(r)\, r}{\sqrt{r^2-R^2}}dr,
\end{equation}
where $\Sigma(R)$ is the projected stellar density. Note that these equations differ in the weighting of the $\beta_a(r)$ parameter for each of the components. 

For the stellar density profile we assume a Plummer model \cite{1911MNRAS..71..460P}
\begin{equation}\label{eqn:plummer}
\nu(r) = \frac{3L}{4\pi R_{HL}^3} \frac{1}{(1+r^2/R_{HL}^2)^{5/2}} \, ,
\end{equation}
for which the projected stellar distribution takes the form
\begin{equation}\label{eqn:plummerProjected}
\Sigma(R) = \frac{L}{\pi R_{HL}^2} \frac{1}{(1+R^2/R_{HL}^2)^{2}} \, 
\end{equation}
where $R_{HL}$ is the projected half light radius. To account for the contribution of the stellar mass to the gravitational potential, we assume a stellar mass-to-light ratio of 1. The actual stellar mass-to-light ratio of Sagittarius is not well constrained, but expectations for a typical dwarf galaxy and previous constraints place it around $\mathcal{O}(1)$ \citep{2001MNRAS.323..529H,2020MNRAS.497.4162V}. We also take the total luminosity of the Sgr core to be $L_{tot}= 10^7L_\odot$, which is consistent with~\citet{Majewski_2003}. 

For the dark matter density profile, we assume a generalized version \citep{1996MNRAS.278..488Z,hernquist1990} of the Navarro-Frenk-White density profile \citep{NFWprofile} given by
\begin{equation} 
\label{eqn:gNFW}
\rho_{DM}(r) = \frac{\rho_s}{(r/r_s)^\gamma [ 1 + (r/r_s)^\alpha]^{(\beta - \gamma) / \alpha}}.
\end{equation}
where $\alpha,\beta,\gamma$ are the power law shape parameters that describe the density profile at small radii $\rho\propto r^{-\gamma}$, the density profile at large radii  $\rho\propto r^{-\beta}$, and the width of the transition between the two regions characterized by $\alpha$.  The scale density and scale radius are $\rho_s$ and $r_s$, respectively. 

The dark matter density profile can be equivalently characterized by a mass and concentration parameter, $M_\triangle=\int_0^{r_\triangle}\rho(r)d^3r$ and $c=r_\triangle/r_s$, where $r_\triangle$ is the radius of the halo. There are several valid choices for defining the halo radius including a virial overdensity, tidal radius, or a fixed threshold of $\triangle=200$ times the mean matter density of the universe. Here we will use the halo parameters $M_{200},c_{200}$. 

In addition to the shape parameters, it will be convenient to describe the halo density profile in terms of the log-slope, defined as $-d\ln{\rho(r)}/d\ln{r}$, and given by
\begin{equation}
    -\frac{d\ln{\rho(r)}}{d\ln{r}} 
    = \gamma - \frac{(\gamma-\beta)(r/r_s)^\alpha}{1 + (r/r_s)^\alpha}
    \label{eq:logslope}.
\end{equation}
This quantity is defined such that for an NFW profile, Equation~\ref{eq:logslope} $\rightarrow 1$ in the center of the halo as $r \rightarrow 0$. On the other hand for a cored Burkert profile, this quantity $\rightarrow 0$ in the center of the halo as $r \rightarrow 0$. The log slope is useful because the shape parameters $\{M,c,\alpha,\beta,\gamma\}$ defining the dark matter profile are often degenerate, and are difficult to constrain from line-of-sight velocity data alone \citep{1982MNRAS.200..361B}. It is possible to constrain more physical parameters such as mass enclosed within the half-light radius $M_{HL}$ and the log slope of the density profile at half-light radius~\citep{2007ApJ...657L...1S,Guerra:2021ppq}.  

\subsection{Bayesian analysis}\label{subsec:bayesAnalysis}

For each of the velocity components, $comp = los, rad, ran$, we define an unbinned likelihood as
\begin{equation}\label{eqn:LOSlikelihood}
\mathcal{L}_{comp} = \prod_{i=1}^N \frac{
\exp\left[ -\frac{1}{2} \frac{\left(u_{i,comp} - \langle u\rangle_{comp}\right)^2}{  \Delta_{comp,i}^2 + \sigma^2_{comp}(R_i)  }\right]
}
{\sqrt{2\pi} \sqrt{ \Delta_{comp,i}^2 + \sigma^2_{comp}(R_i) } } .
\end{equation}
with the product being over the number of stars. The dispersion is the sum of the projected, intrinsic dispersion from the Jeans equation, and the measurement uncertainty of a given star, $\Delta_{comp,\imath}$. The quantity $\langle u\rangle_{comp}$ represents the mean velocity of the stars in each component direction, after COM motion has been subtracted. This should be zero, or consistent with zero, in the case with no overall rotation or tidal stripping  as assumed in Section \ref{subsec:jeansAnalysis}. A departure from zero would indicate that this assumption is not valid, and would not hold at larger radii where tidal stripping becomes important. It is for this reason that we have restricted our data to small radii. For the case of the analysis of all three components, the likelihood is $L = \prod_{comp} L_{comp}$. 

To estimate the likelihood, we use the Bayesian inference tool {\tt MultiNest} through the python interface {\tt PyMultiNest} \citep{MultiNestDocs,pymultinestDoc}. {\tt MultiNest} operates by sampling $N$ points from the input prior space, then discarding the lowest likelihood $L_0$ point. It is replaced by a new point with likelihood $L_1$ if $L_1>L_0$, and the prior volume is reduced. 
We sample over the following flat priors, in logarithmic space for velocity anisotropy, halo mass, and concentration parameters and linear space for generalized NFW profile shape parameters $\alpha, \beta, \gamma$.
\begin{equation}\label{eqn:priors}
\begin{aligned} 
 -1 \leq -\log_{10} (1-\beta_a) \leq +1, \\
 \log_{10}(5\times 10^7) \leq \log_{10} (M_{200} / M_\odot) \leq \log_{10}(5\times 10^9) , \\
 \log_{10}(2) \leq \log_{10} (c_{200}) \leq \log_{10}(30) , \\
 0.5 \leq \alpha \leq 10, \\
 3\leq \beta \leq 20, \\
 0 \leq \gamma \leq 3. \nonumber
\end{aligned}
\end{equation}

We tested larger prior ranges, including an increased $M_{200}$ range and allowing for negative $\gamma$, but saw no change in the resulting posteriors.
The last important input is a value of projected half-light radius. Determination of the half-light radius for Sgr is complicated because the core blends into the tidal tails. Further, the estimate of the tidal radius depends on the stellar population considered. Previous estimates of the half-light radius range from $R_{HL}\simeq2.6$ kpc in \citet{McConnachie_2012} and $R_{HL}\sim1.1$ kpc in \citet{2020MNRAS.495.4124F}. This latter value is also consistent with the determination from~\citet{2021ApJ...908..244D}. For our analysis below, we use a half-light radius fit from our data that is consistent with the smaller of the two from previous literature.


\section{Data}\label{sec:data}

In this section, we present our analysis of the Gaia data, and how member stars for Sgr are identified. We then discuss how uncertainties on the distance to each star are included in the analysis.  

We use data from the Gaia mission~\citep{2016A&A...595A...1G}, and specifically from the third data release (DR3)~\citep{2023A&A...674A...1G}. In order to select stars associated with the core of Sgr, we first query all stars approximately within a 5.7 degree radius around the center of the galaxy, $(\alpha_0, \delta_0)=(283.83^{\circ}, -30.55^{\circ})$. This corresponds to the larger of the two measured values of the half-light radius discussed above, projected onto the sky, and is defined by the region $(277^{\circ},-37^{\circ})<(\alpha, \delta)<(290^{\circ},-24^{\circ})$. We then make a cut to only include stars that have parallax values $\Pi < 1$ mas, which removes nearby stars in the MW disk. 
Finally, we take the subset of these stars that are bright enough to have measured Gaia line-of-sight velocities. These cuts return a sample of $25,777,561$ stars that have a wide distribution in proper motion space. See Appendix~\ref{app:queries} for the specific query used to select this sample. 

\begin{figure*}
\begin{subfigure}{0.54\textwidth}
    \includegraphics[width=\textwidth]{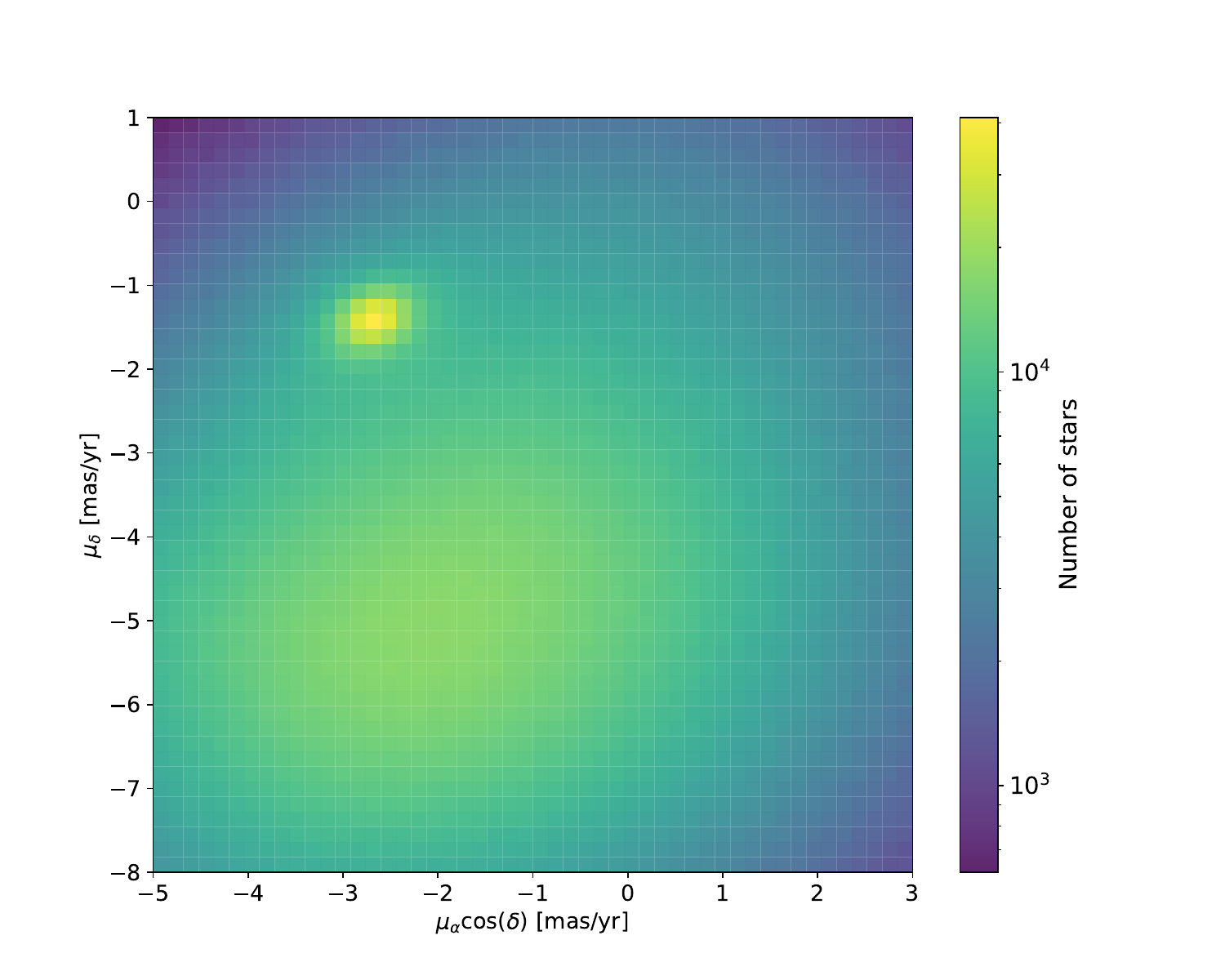}
\end{subfigure}
\caption{Proper motion space histogram of all stars returned by the Appendix \ref{app:queries} query, with logarithmic normalization. }
\label{fig:PMspace}
\end{figure*}

\begin{figure*}
\includegraphics[width=0.9\textwidth]{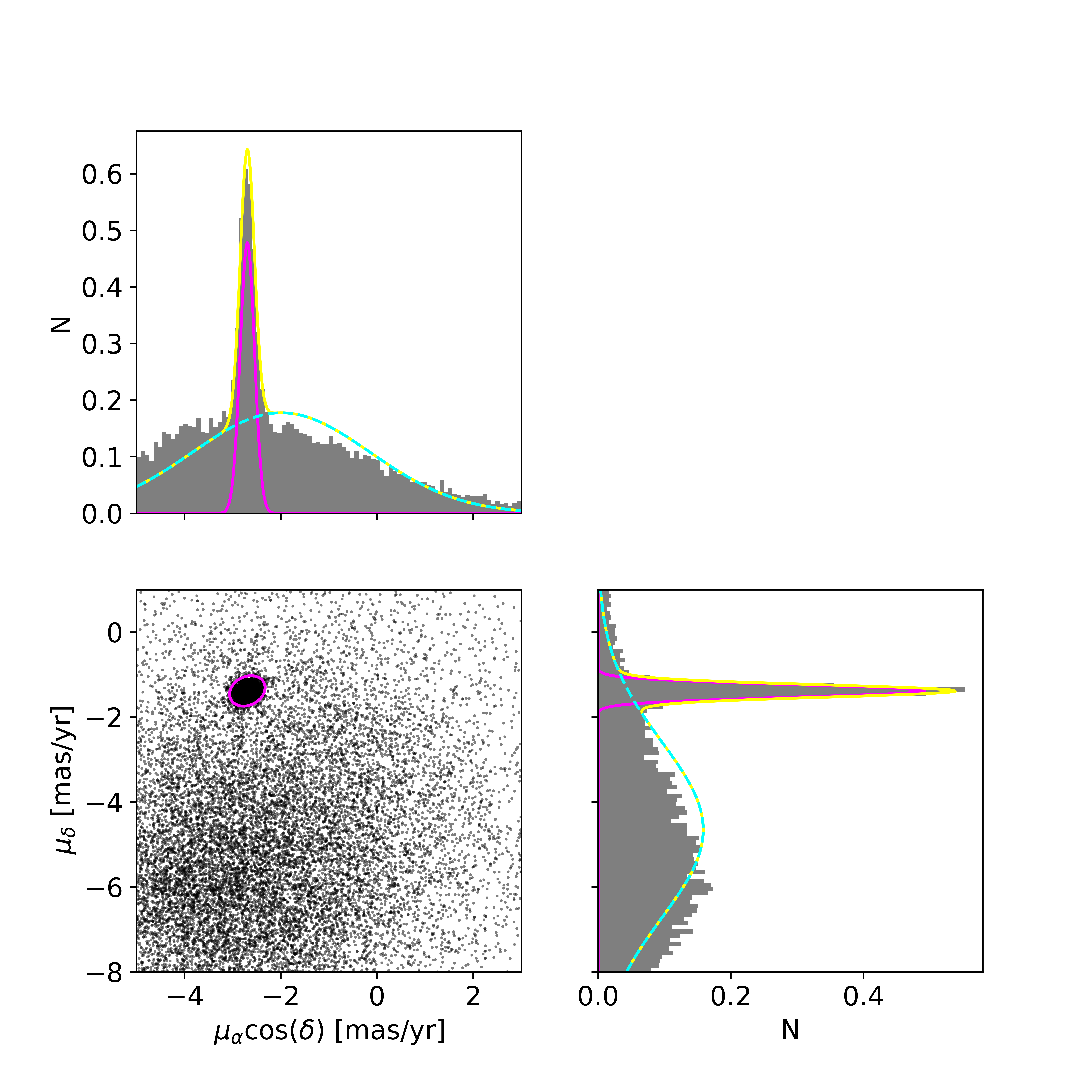}
\caption{ Bottom left panel shows stars in proper motion space after parallax over error and radial velocity cuts described in the text. The pink ellipse represents the Gaussian mixture model fit 2$\sigma$ region of Sagittarius stars. The top left and bottom right panels show the projected PDFs plotted over a normalized histogram of the stars in $\mu_{\alpha*}, \mu\delta$ respectively. The solid pink line corresponds to the Sgr component, the cyan dotted line to the field star component, and the solid yellow line to the sum of both components. 
}
\label{fig:projectedGMMs}
\end{figure*}

Figure~\ref{fig:PMspace} shows a 2D histogram of all stars in proper motion space returned by the original query with logarithmic normalization. The bottom left panel of Figure~\ref{fig:projectedGMMs} shows a scatter plot in proper motion space of stars that pass the cuts described above. The Sgr region of proper motion space is identified by fitting a two component Gaussian mixture model to the selected stars in this region. Sgr member stars are identified as within the $2\sigma$ ellipse of the Gaussian Sgr component.
The ellipse is given by means $\bm{\mu}\simeq(-2.695, -1.384)$[mas/yr], $\bm{\sigma}\simeq(0.129, 0.160)$[mas/yr], semi major and minor axes=(0.640, 0.797)[mas/yr], and a counterclockwise rotation of $\theta=-0.861$ rad. For a more detailed discussion on similar methods for selecting member stars in dSphs, and the systematics associated with foreground subtraction, see~\citet{2022ApJ...940..136P}. The bottom left panel of Figure~\ref{fig:projectedGMMs} shows the resulting $2\sigma$ ellipse that contains the Sgr component. The top left and bottom right panels show the projected PDFs plotted over a normalized histogram of the stars in $\mu_{\alpha*}, \mu_\delta$ respectively, with lines corresponding to the Sgr and field components.

To further minimize the number of field stars that may be confused with Sgr member stars, a line-of-sight velocity cut of $|v_{LOS}|<40$ km/s is applied after the coordinate transform described in Section~\ref{subsec:coordTrans}, leaving 2729 candidate stars. Note that this is after subtracting the center of mass motion, so this is cutting stars $>40$ km/s from the $\sim143$ km/s mean. The $|v_{LOS}|<40$ km/s cut was chosen to roughly remove outliers. Assuming Gaussian errors, this corresponds to an approximate 2.5$\sigma$ clip of outliers, and a conservative cut for removing high velocity outliers and interlopers from the sample. Note that after the coordinate transformation, the mean velocity in each of the different component directions vanishes. We then fit a projected Plummer distribution to the resulting member star sample, and find the projected half-light radius to be $2.601^\circ \pm  0.002^\circ$, or $R_{HL}\approx1.18$ kpc with a central distance of $D_0=25.951$kpc \citep{McConnachie_2012}. From this point on any use of the half-light radius will be this value. This is consistent with estimates discussed above that have been obtained from Gaia data. 

\begin{figure*}
\includegraphics[width=0.9\textwidth]{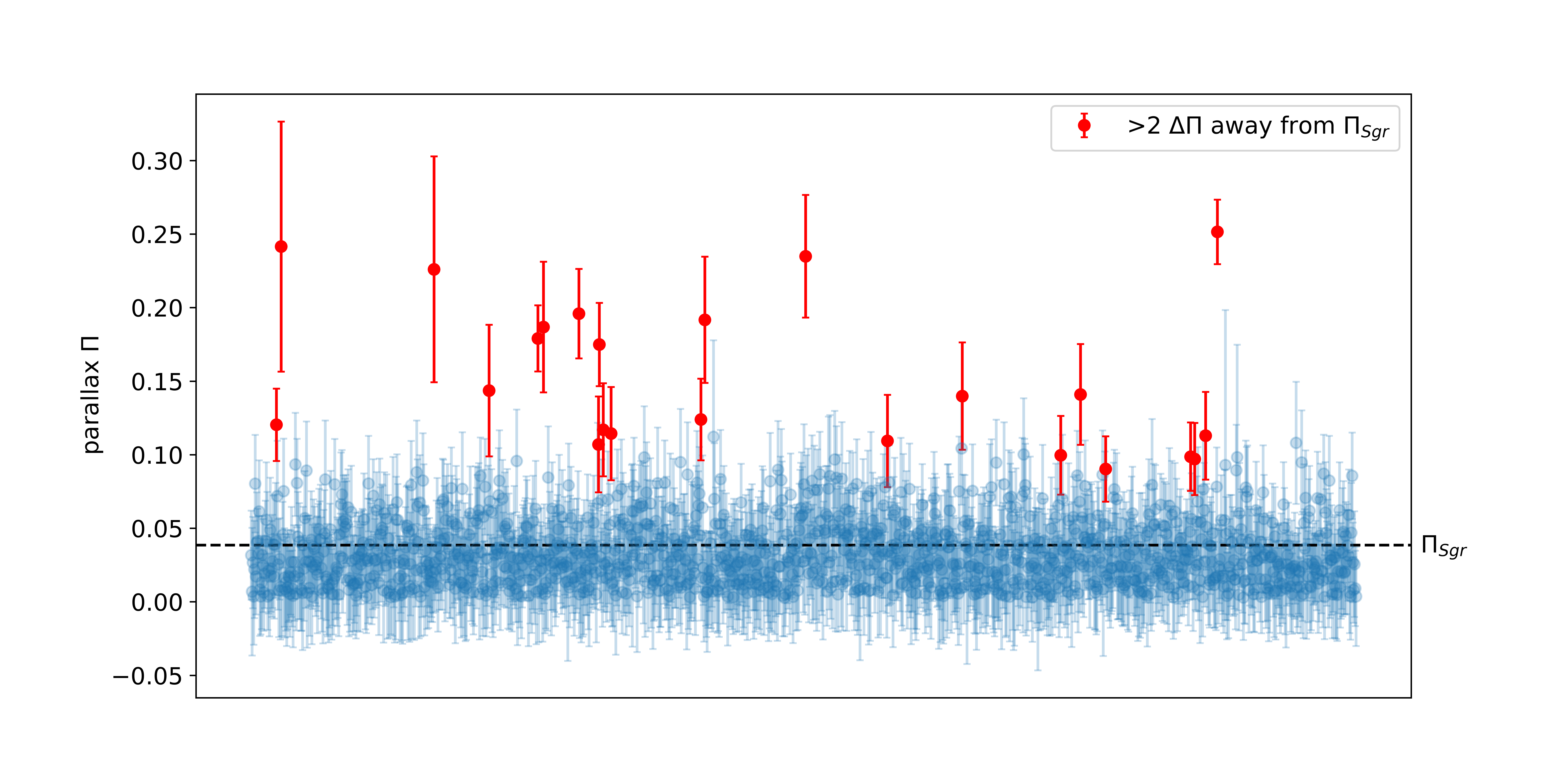}
\caption{Parallaxes of the Sgr core candidates. The x axis is for readability, and does not represent a physical quantity. Error bars are the reported parallax error $\Delta\Pi$ from Gaia. Red highlighted points are those that are greater than $2\Delta\Pi$ away from $\Pi_{Sgr}$, which are subsequently removed from the candidate list. 
}
\label{fig:parallaxes}
\end{figure*}

\begin{figure*}
\includegraphics[width=0.9\textwidth]{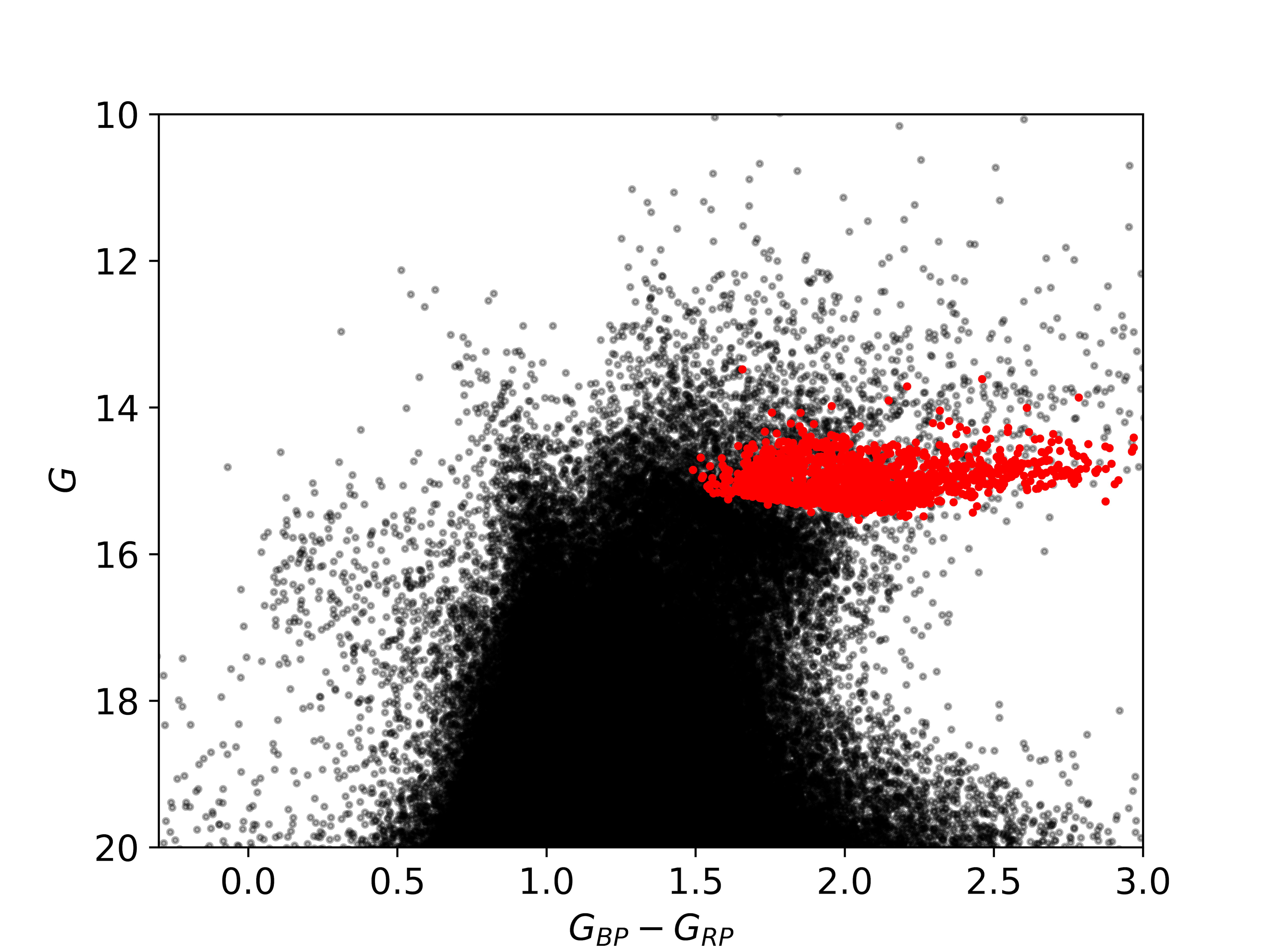}
\caption{Color magnitude diagram of the Sagittarius candidates. Black points are all stars from the queried data in the region of Sagittarius' proper motion space, and red points are the final candidates. 
}
\label{fig:CMD}
\end{figure*}

In a conservative effort to remove any possible effects of tidal disruption on our member sample, our final set of Sgr candidates includes only stars within this measured half-light radius of $2.6^\circ$. This results in 1404 Sgr candidates, which are shown in Figure~\ref{fig:parallaxes}. The 23 red points are those that are more than two times their error away from the expected mean parallax of Sgr, $\Pi_{Sgr}\sim 1/D_0$, and are removed from the candidate list. The minimum membership probability of stars within this final sample is $\sim0.85$ as calculated by the Gaussian mixture model. A color magnitude diagram of this final selection of 1381 stars is shown in Figure~\ref{fig:CMD}, and a list of Gaia DR3 source\_ids, with their associated membership probabilities as calculated by the Gaussian mixture model, for final selected candidates is published in the auxiliary file {\tt Sgr\_candidates.txt}.

Measurements of parallaxes for stars at the distance of Sgr are neither precise nor accurate enough to utilize in our Jeans analysis below. Indeed, only one star in our sample has a parallax measured to less than 10\%, i.e. $\Delta\Pi/\Pi<0.1$. In light of this, to assign distances to the stars we generate two mock catalogs; one catalog has an error on each distance ``measurement'' of $\Delta_D=0.1$ kpc and one has $\Delta_D=1$ kpc. To generate mock distances for each individual star, we sample from a normal distribution centered at the distance to the center of Sgr-- 25.95 kpc~\citep{McConnachie_2012, 2021ApJ...908..244D}-- and standard deviation of 1 kpc.

With these mean distances sampled for each star, to propagate errors, we sample values of $(\alpha, \delta, D, \mu_{\alpha*}, \mu_\delta, \vlos)$ 10,000 times from normal distributions centered at their measurement value with standard deviations of their measurement error. The coordinate transform is then performed for each of the 10,000 samples and the standard deviation of the resulting $(\rho, \phi_\ast, \vlos, \vrad,\vtan)$ distributions is used as errors on those transformed parameters. For simplicity we have not included the $\alpha, \delta, \mu_{\alpha*}, \mu_\delta$ systematic errors and assumed them to be uncorrelated although both are present in Gaia measurements \citep{2021A&A...649A...2L}. No noticeable difference is found in RAD+TAN posteriors after applying the Gaia systematics correction for correlated proper motions found in \citet{2021MNRAS.505.5978V}. The details of that analysis are given in Appendix~\ref{app:extraanalysis}.
 
\begin{figure*}
    \includegraphics[width=0.7\textwidth]{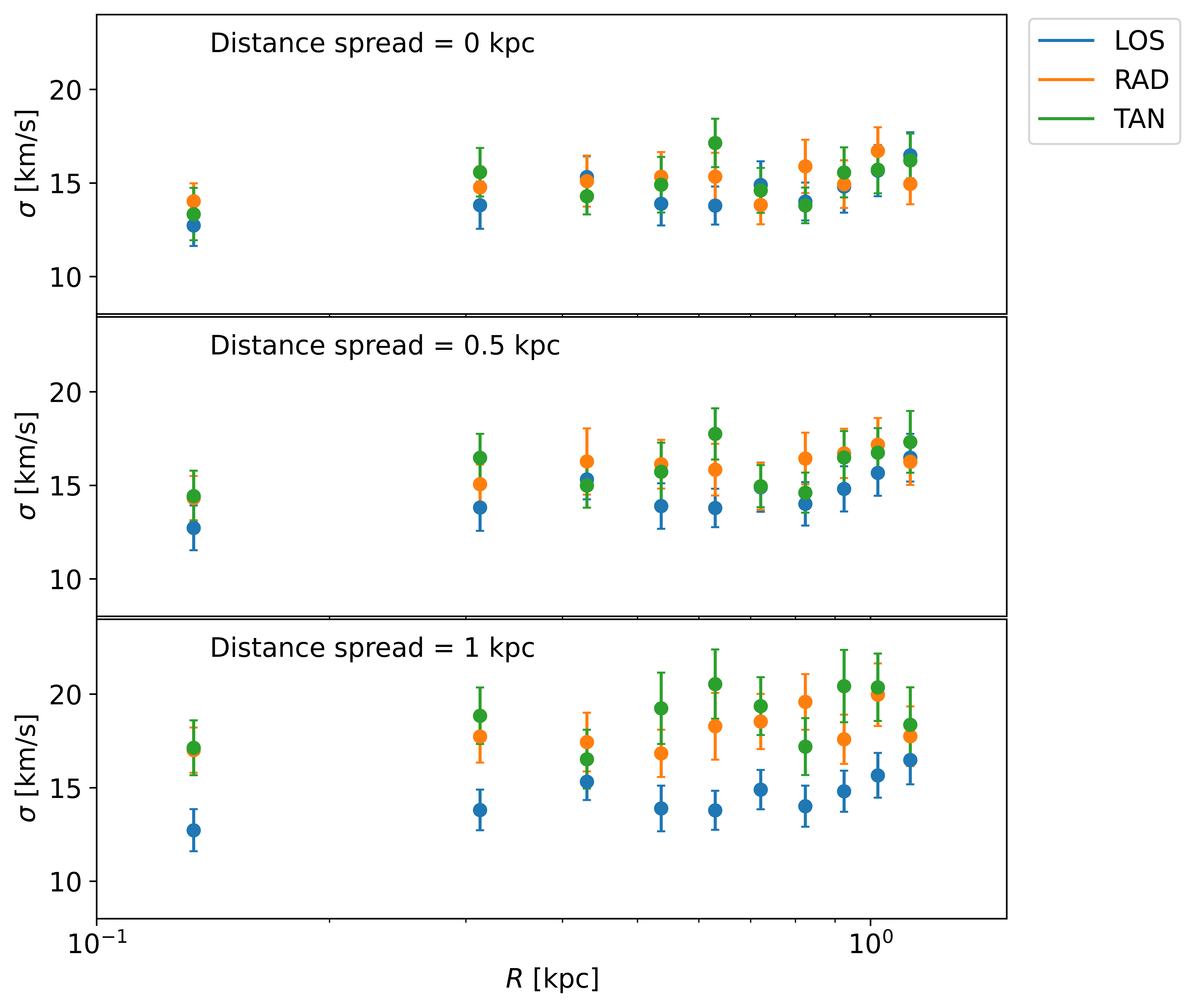}
\caption{Binned velocity dispersions of the sample stars with three sets of mock distances out to the half light radius. Top panel shows all stars at the same distance $D_0$, middle panel has randomly selected distances from a Gaussian distribution with standard deviation 0.5 kpc, and bottom panel has randomly selected distances from a Gaussian distribution with standard deviation 1 kpc. Bins are constructed to contain approximately equal numbers of stars, and error bars are estimated with the jackknife method. 
}
\label{fig:binneddisp_changeDistSpread}
\end{figure*}

Figure~\ref{fig:binneddisp} shows the resulting velocity dispersion for the different velocity components. Beyond the measured half-light radius, there is an increase in the line of sight velocity dispersion, the only component not influenced by mock distances. This inflation may be a result of contamination from stars in the tidal tails, or possibly a small number interloping stars that are not members of Sgr. This increase beyond the half-light radius justifies our sample cut which we use in the Jeans analysis below. Note that the error bars shown in Figure~\ref{fig:binneddisp} are jackknife resampling estimates made from the stars' $v_{comp}$ value in a given bin, not propagated from the errors on those velocities, and as such will be the same for both $\Delta_D=0.1$ kpc and $\Delta_D=1$ kpc cases. However, changing the distribution of mock distances will change the velocity dispersion for $rad$ and $tan$ components. Figure \ref{fig:binneddisp_changeDistSpread} shows the effect of setting the distance of all stars to $D=D_0$, randomly selecting distances from a Gaussian distribution with standard deviation 0.5 kpc, and randomly selecting distances from a Gaussian distribution with standard deviation 1 kpc (as used in our analyzed sample).

\begin{figure*}
\begin{subfigure}{0.7\textwidth}
    \includegraphics[width=\textwidth]{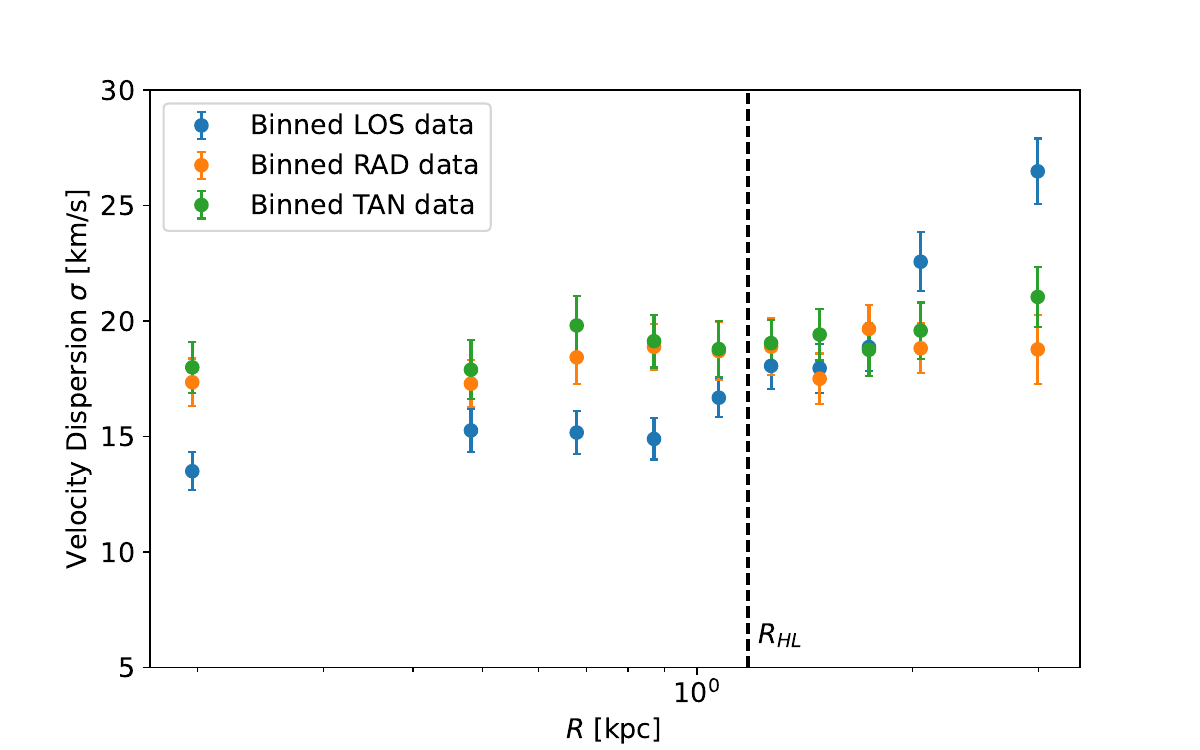}
\end{subfigure}
\caption{Binned velocity dispersion of our sample with mock distances showing the line of sight, radial, and tangential components. Bins are constructed to contain approximately equal numbers of stars, and error bars are estimated with the jackknife method. Note that the jackknife error bars remain unchanged for $\Delta_D=0.1$ kpc or $\Delta_D=1$ kpc. Binning is for display only; unbinned data and errors are used in the analysis. 
}
\label{fig:binneddisp}
\end{figure*}

\begin{figure*}
\includegraphics[scale=0.75]{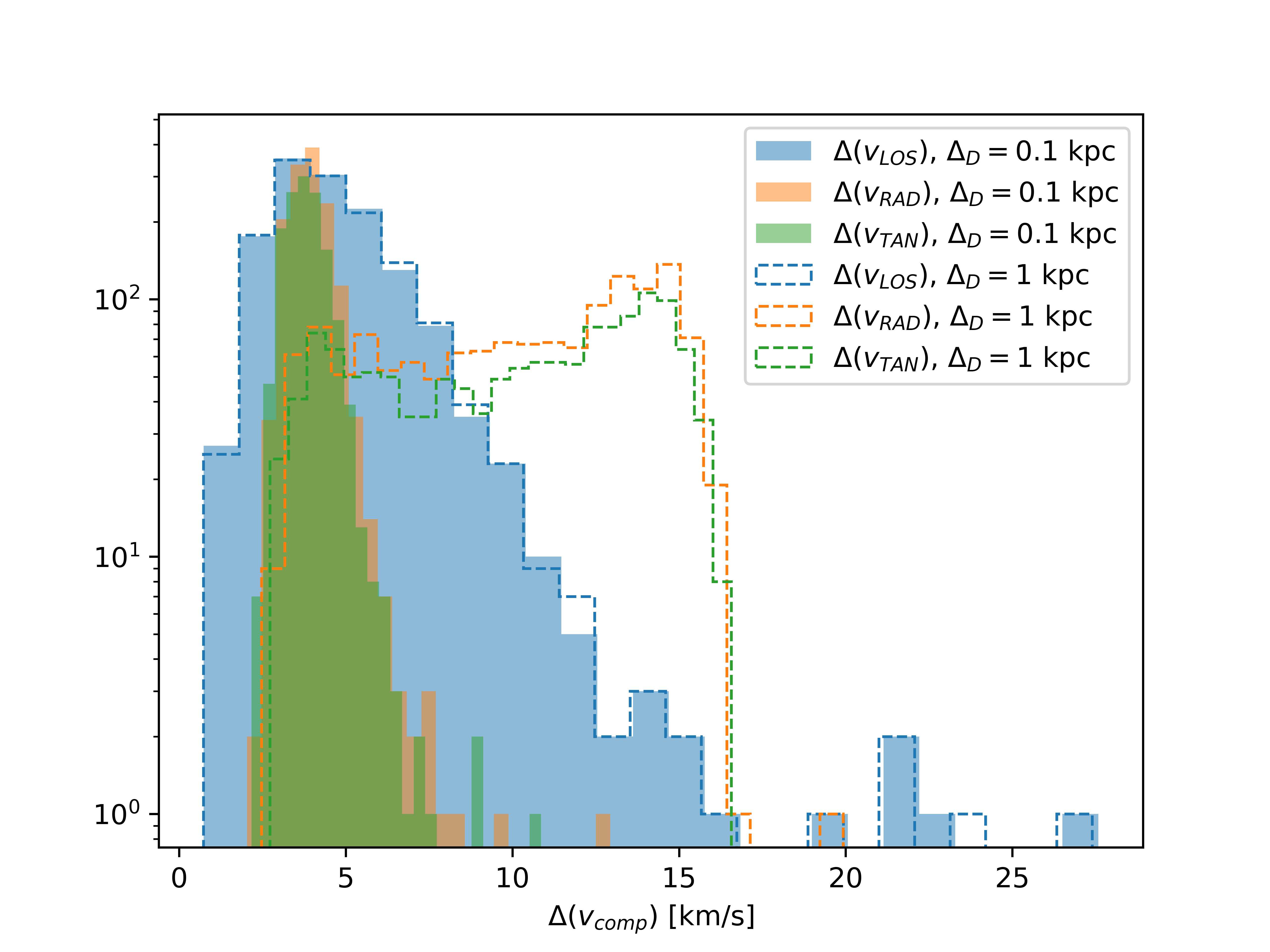}
\caption{A comparison of the propagated errors $\Delta(v)$ in line-of-sight (blue), radial (orange), and tangential (green) velocities in the case of $\Delta_D=0.1$ kpc distance errors (solid histogram) and $\Delta_D=1$ kpc distance errors (dashed histogram).}
\label{fig:distErrProp}
\end{figure*}

Figure \ref{fig:distErrProp} shows the effect of larger distance errors $\Delta_D$ on the resulting $\vlos, \vrad,\vtan$ errors for Sgr candidates. The radial and tangential components are affected, while the line-of-sight component, measured directly and with no dependence on stellar distance in the coordinate transformation, is not. As one might expect, the $rad$ and $tan$ velocity errors increase as $\Delta_D$ increases. Since there is larger uncertainty per star, this has the effect of weighting the $rad$ and $tan$ dispersion components less in the Bayesian analysis. 

\section{Results}\label{sec:results}

We perform separate {\tt MultiNest} runs for the {\it rad}+{\it tan} combined (proper motion components) and {\it los} component for both the $\Delta_D=0.1$ kpc and $\Delta_D=1$ kpc datasets. For each dataset we also perform a run in which the likelihoods for all three velocity components are combined (a 6D analysis, abbreviated as Sgr All). As described above, in our runs we vary the shape parameters of the density profile, along with the scale mass, the scale concentration, and the velocity anisotropy, the latter of which is assumed to be constant over $R$. We then convert these quantities so that the reported posteriors are the parameters of velocity anisotropy $\log_{10}(1-\beta_a)$, dark matter mass contained within the half light radius $\log_{10}(M_{HL}/M_\odot)$, concentration $\log_{10}(c)$, and log slope at $R_{HL}$.

\begin{figure*}
\begin{subfigure}{0.6\textwidth}
    \includegraphics[width=\textwidth]{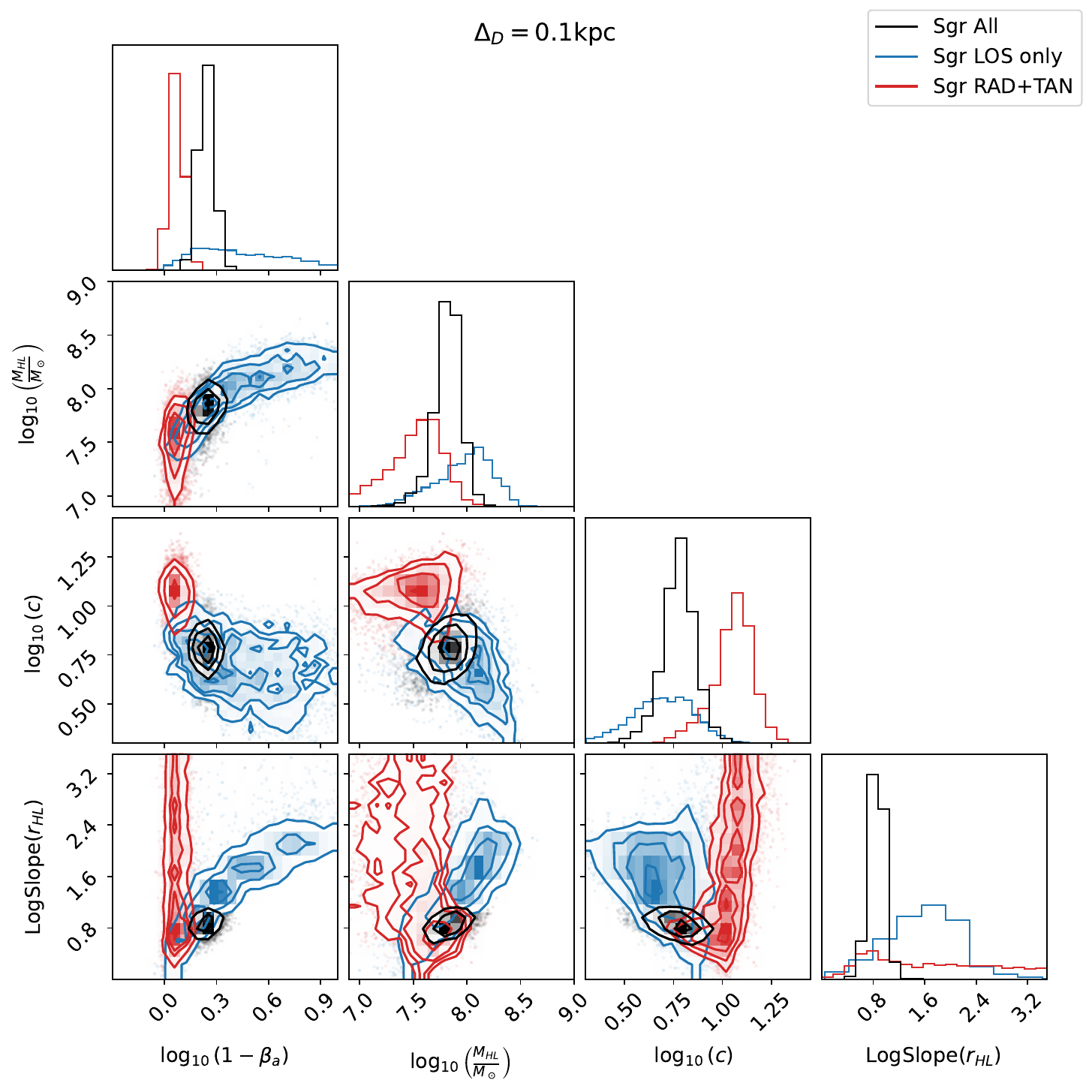}
\end{subfigure}

\begin{subfigure}{0.6\textwidth}
    \includegraphics[width=\textwidth]{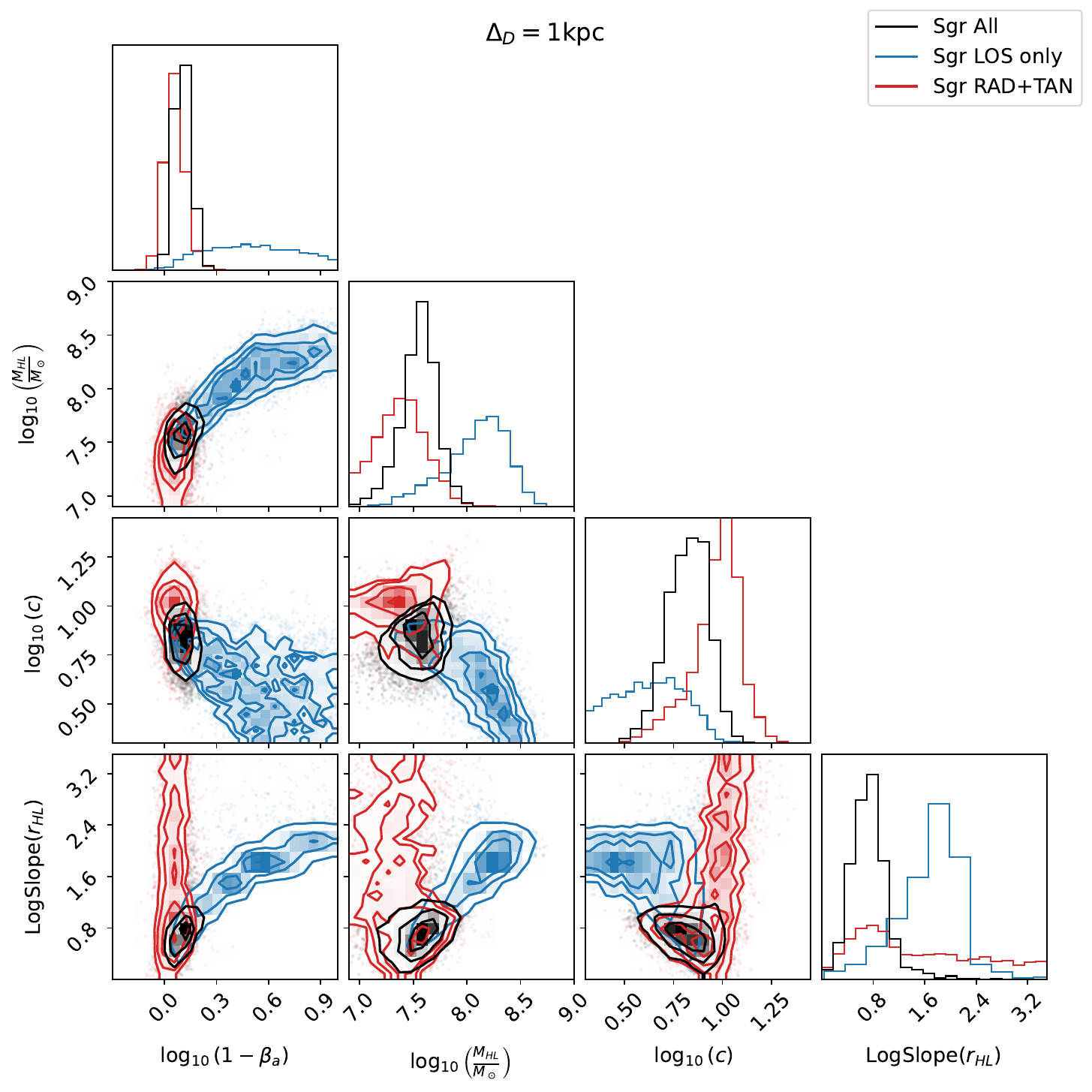}
\end{subfigure}
\caption{Top corner plot shows posteriors for half light radius of 1.18 kpc and $\Delta_D=0.1$ kpc. Bottom corner plot shows posteriors for half light radius of of 1.18 kpc and $\Delta_D=1$ kpc. Black contours correspond to a 6D (all components) analysis, blue contours correspond to a line of sight velocity analysis only, and red contours correspond to a combined radial and tangential velocity analysis. In both corner plots, the parameters are velocity anisotropy in the form $\log_{10}(1-\beta_a)$, mass contained within the half light radius $\log_{10}(M_{HL}/M_\odot)$, concentration $\log_{10}(c)$, and log slope at the half light radius LogSlope($r_{HL}$).}
\label{fig:compareLOS_RADTAN}
\end{figure*}

The resulting posteriors for each of the different runs are shown in Figure~\ref{fig:compareLOS_RADTAN}. In both the $\Delta_D=0.1$ kpc and $\Delta_D=1$ kpc cases, the 6D analyses show good agreement with LOS posteriors. As expected, there is a strong degeneracy between the log slope and the velocity anisotropy, which is particularly evident for the LOS-only data set. Also as expected, the mass within the half-light radius is well constrained by all of the data sets, and there is good agreement between the posteriors in all cases. The mass within the half-light radius is consistent between the two distance uncertainties. The measurement of the velocity anisotropy is also significantly improved with the inclusion of all three velocity components. We again emphasize that for the {\it los} data set the mean and the uncertainties on the posteriors reflect a measurement of the corresponding quantity. On the other hand, for the {\it rad} and {\it tan} combined run, our analysis should be taken as a robust estimate of the expected measurement uncertainties, while the mean estimated quantities may be biased by the model for the distance uncertainties.

As indicated in Figure~\ref{fig:compareLOS_RADTAN}, there are relatively weak constraints on the log-slope for an analysis which uses only any one of the three individual velocity components. However, there is significant improvement when using a likelihood that combines all three velocity components. In particular, in the analysis of the combined three data sets, the degeneracy between the log slope and the velocity anisotropy is broken, and a mild anisotropy is identified. This can also be seen by comparing the {\it rad} and {\it tan} run and the {\it los} only run posteriors. The addition of proper motions breaks the log slope and velocity anisotropy degeneracy. In the LOS only case, which uses only real data, anisotropy is measured to be $\beta_a=-2.24\pm1.99$, implying a tangentially biased system with large uncertainty. This is consistent with previous measurements of a tangentially biased Sgr \citep{2013ApJ...777L..13M}. The log slope at $R_{HL}$ is $1.62 \pm 0.69$. These errors improve in the full 6D analysis; however it is important to note that since mock distance values are used, this is an estimate of the constraining power of a 6D analysis rather than a measurement. For the $\Delta_D=0.1$ kpc dataset, anisotropy is measured to be $\beta_a=-0.73\pm0.18$ and log slope at $R_{HL}$ is $0.85\pm0.13$. For the $\Delta_D=1$ kpc dataset, anisotropy is measured to be $\beta_a=-0.28\pm 0.14$ and log slope at $R_{HL}$ is $0.74\pm 0.31$.

\begin{figure*}
\begin{subfigure}{0.4\textwidth}
    \includegraphics[width=\textwidth]{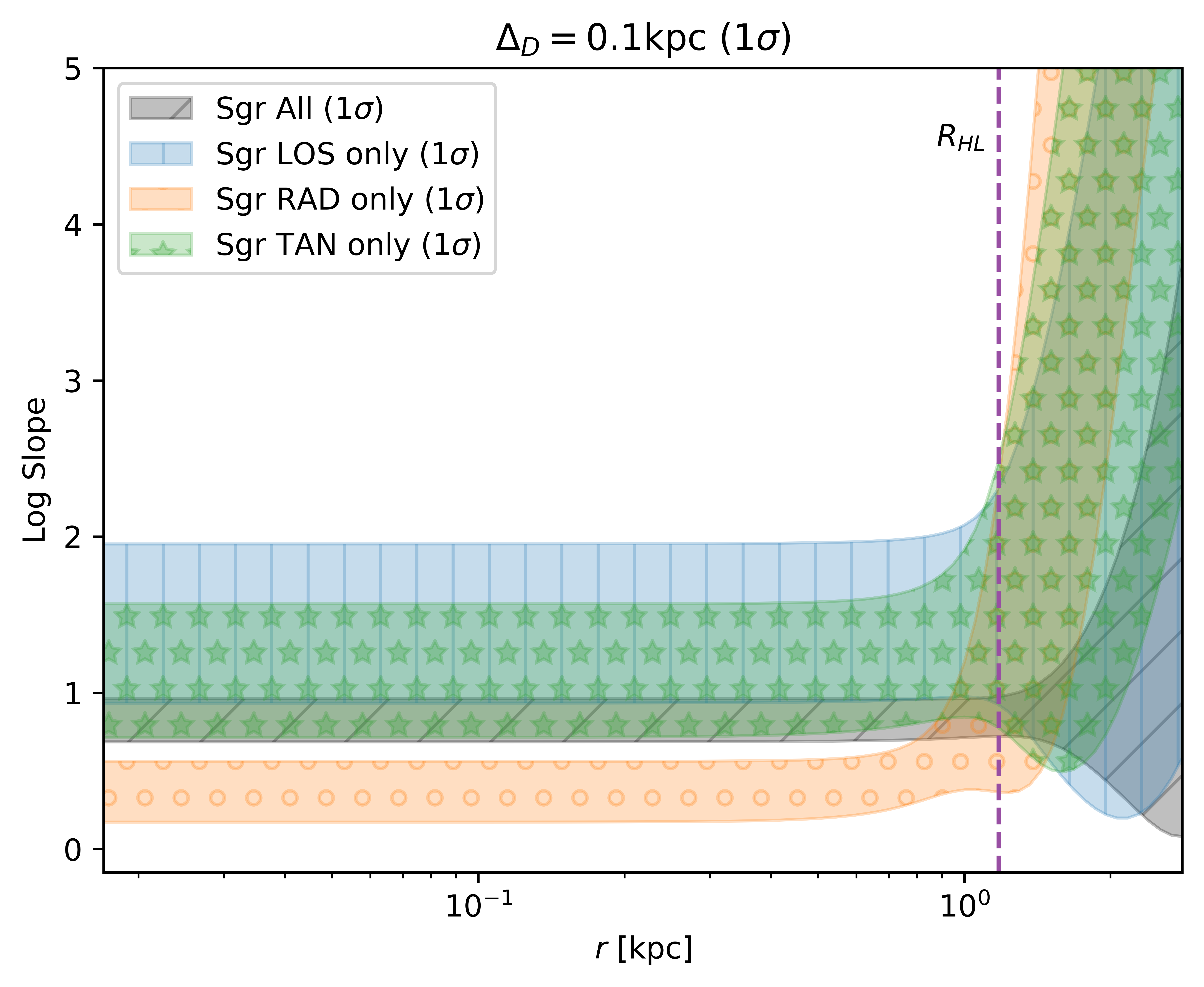}
\end{subfigure}
\begin{subfigure}{0.4\textwidth}
    \includegraphics[width=\textwidth]{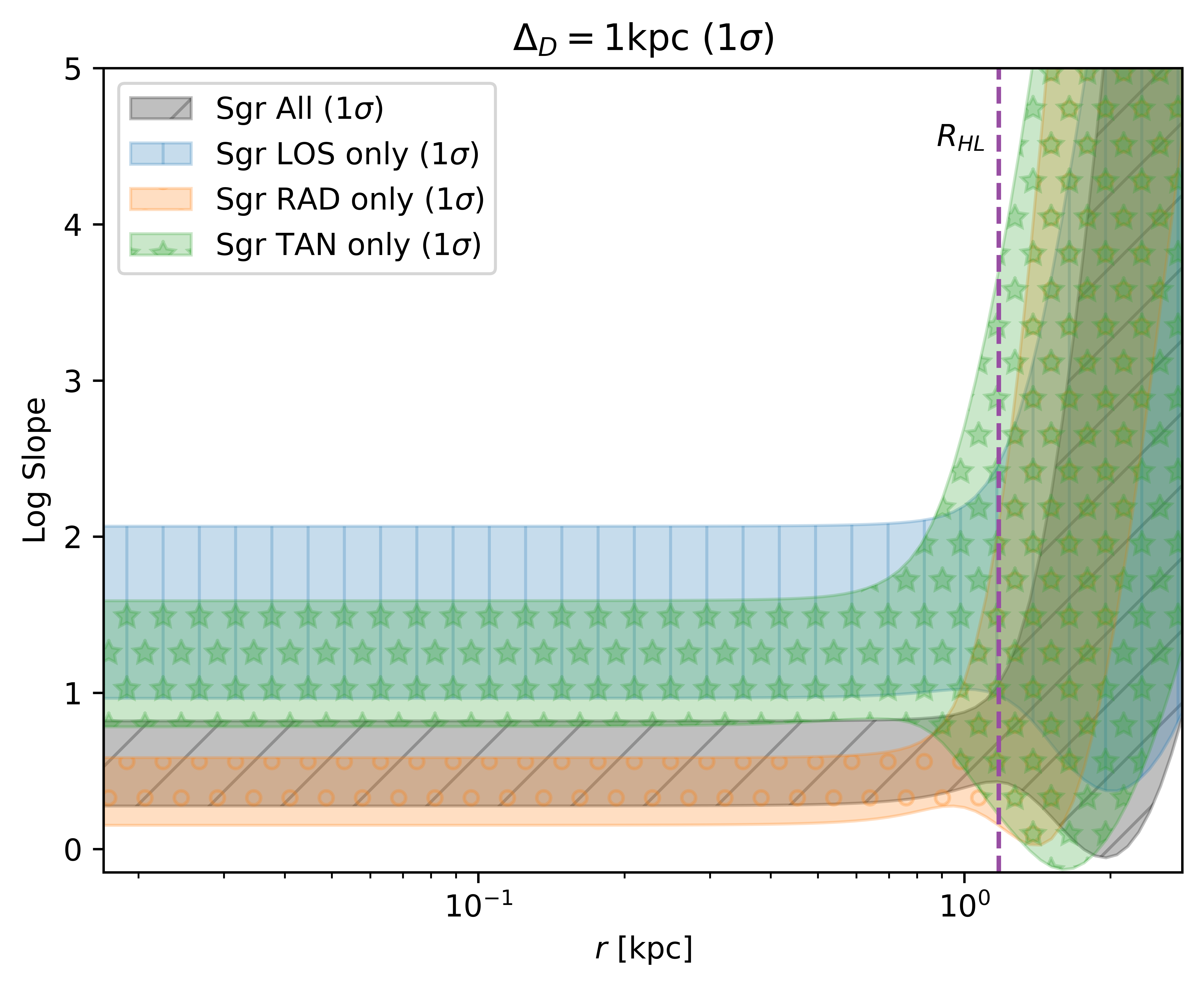}
\end{subfigure}
\begin{subfigure}{0.4\textwidth}
    \includegraphics[width=\textwidth]{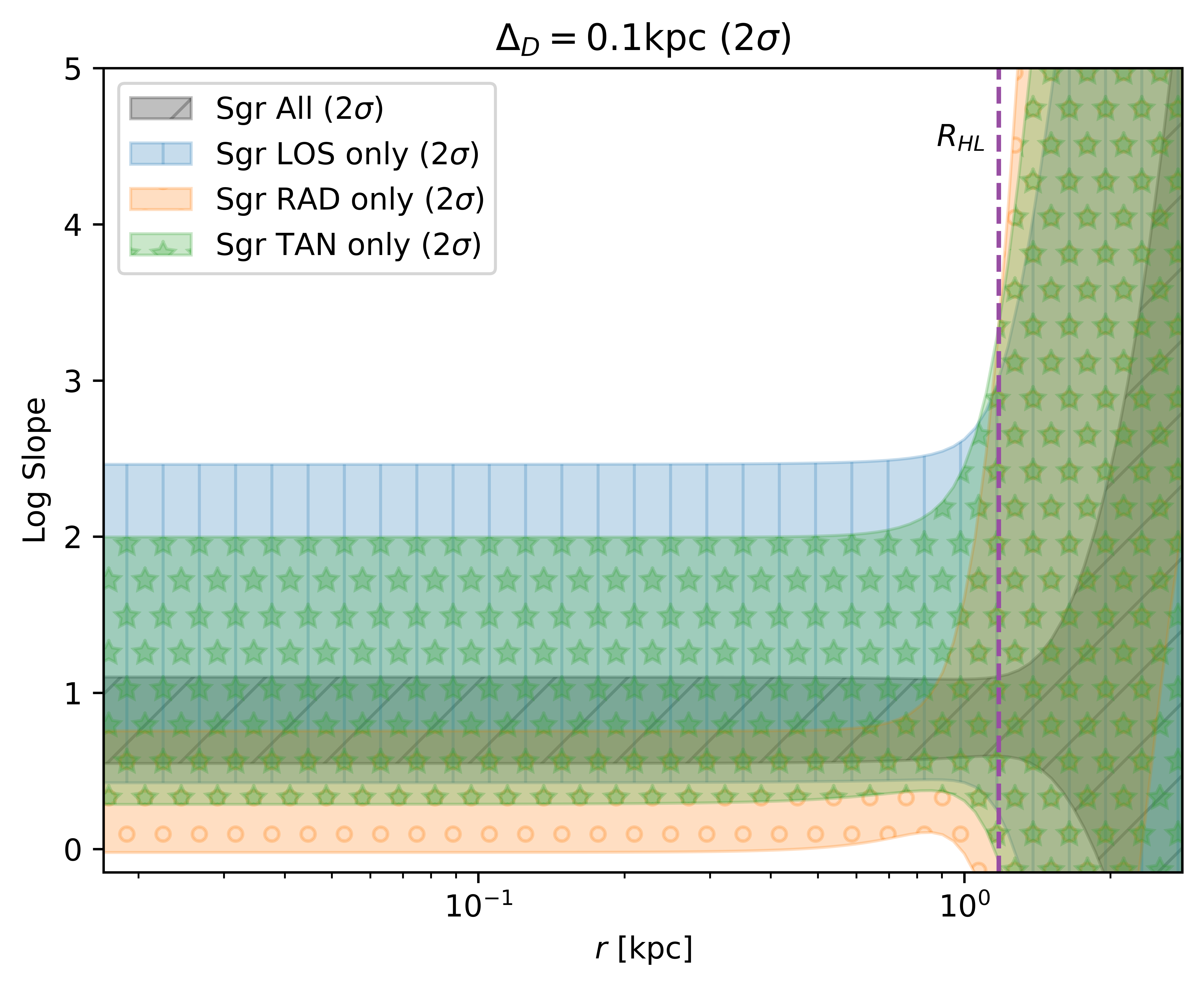}
\end{subfigure}
\begin{subfigure}{0.4\textwidth}
    \includegraphics[width=\textwidth]{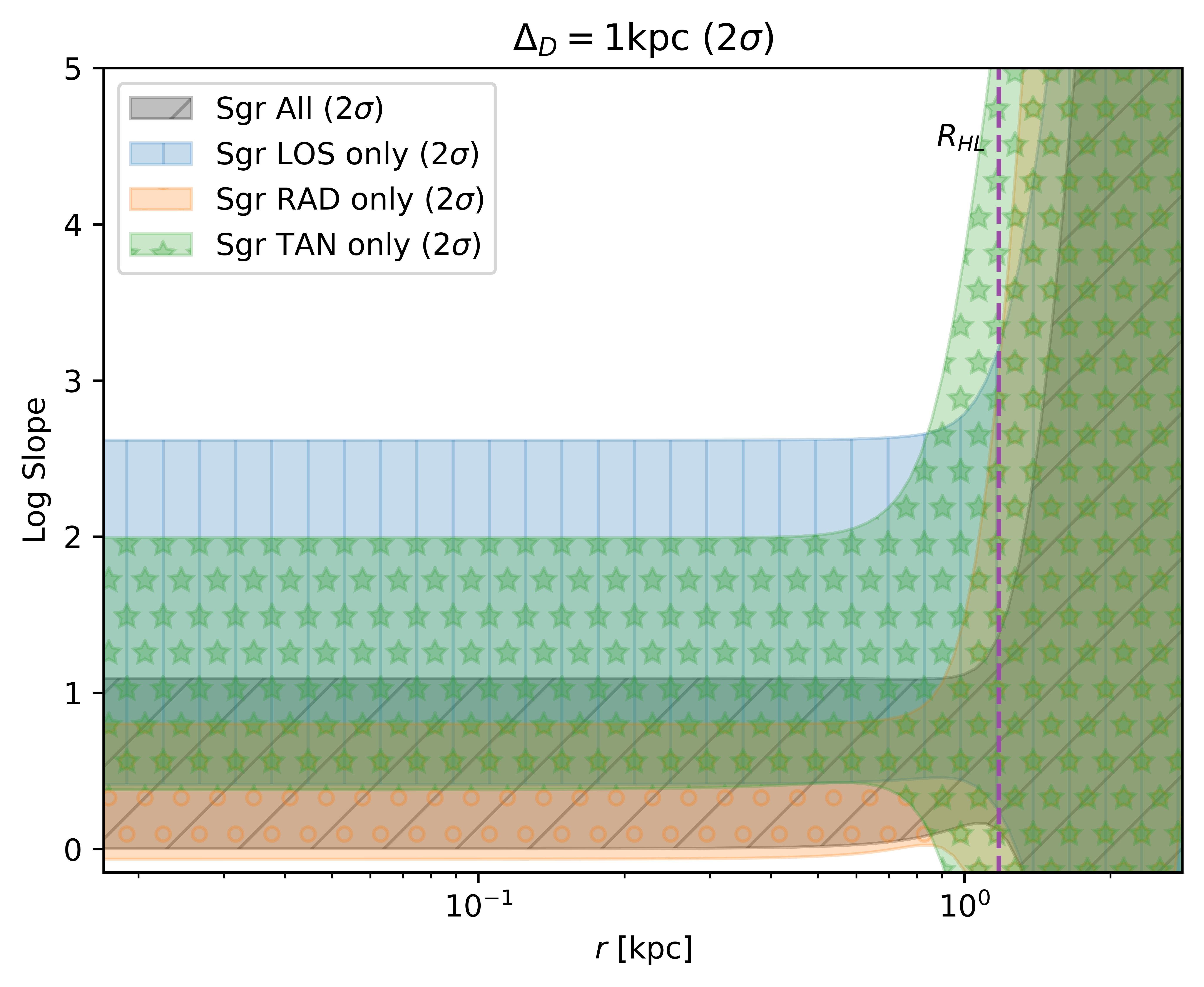}
\end{subfigure}
\caption{Log slope as a function of spherical coordinates radius calculated from posteriors, for $\Delta_D=0.1$ kpc (left column) and $\Delta_D=1$ kpc (right column). Colored bands show the range of $1\sigma$ (top row) and $2\sigma$ (bottom row) value of log slope at a given $r$. Black with diagonal hatching represents the 6D (all components) Jeans analysis, blue with vertical hatching represents a line of sight velocity only analysis, orange with circle hatching represents a radial velocity only analysis, and green with star hatching represents a tangential velocity only analysis.
The purple vertical dotted line is at the projected half light radius $R_{HL}=1.18$ kpc. 
}
\label{fig:logslope}
\end{figure*}

Figure~\ref{fig:logslope} shows the uncertainty on the log-slope as a function of radius, for both of the $\Delta_D$ datasets. Across all radii, the full likelihood with all velocity components provides the best constraints on the log-slope. It is evident for the smaller distance uncertainty case that the best uncertainty on the log-slope is obtained just below the half-light radius. While the constraint on the log-slope depends on radius, even with the full three-component likelihood the asymptotic central slope for an NFW profile is still allowed. From the right-hand panel, we see that the reconstruction of the log-slope degrades for the case of larger distance uncertainties. This result clearly emphasizes that distance uncertainties will be important for accurately reconstructing the log slope, and possibly distinguishing between and NFW-based model or its alternatives. 

\section{Conclusion}\label{sec:conclusion}

\par In this paper we have performed an analysis of the kinematics in the Sagittarius dwarf spheroidal core, considering all three velocity components, with particular focus on the effect of errors on stellar distance measurements. We have identified a sample of bright stars from Gaia DR3 which have full 3D velocity information. Our analysis represents the first kinematic analysis of Sagittarius with self-consistent kinematics derived from a selected sample of stars. Using the line-of-sight velocity component combined with an equilibrium-based Jeans model, we obtain a measurement of the log-slope of the density profile at $R_{HL}=2.6^\circ$--- Log Slope($R_{HL}$)=$1.62 \pm  0.69$--- and the velocity anisotropy $\beta_a=-2.24\pm1.99$. 

\par For the two transverse velocity components, we pay particular attention to how the uncertainties in distances propagate through to values of kinematic quantities. We show that uncertainties on the distance inflate the uncertainties for radial and tangential velocities. Additionally, an increased spread in the distance values themselves inflate the velocity dispersion of the radial and tangential components. Thus it is very important to have both precise and accurate measurements of stellar distances for a fully six dimensional Jeans analysis. This is reflected in the velocity anisotropy and log slope posteriors where we see a smaller error for lower distance measurement errors: for the case of smaller errors on distance measurements, $(\beta_a,$Log Slope($r_{HL}$))=$(-0.73,0.85)\pm(0.18, 0.13)$ and for the case of larger errors on distance measurements $(\beta_a,$ Log Slope($r_{HL}$))=$(-0.28, 0.74)\pm(0.14, 0.31)$. 

\par We anticipate that analyses along the lines that we consider, which include real distance measurements for stars, will reduce the uncertainties on measurements of the log-slope of the density profile and the velocity anisotropy in a similar fashion to our mock distances case. We predict that uncertainties on velocity anisotropy and log-slope can be reduced by an order of magnitude and a half order of magnitude respectively. These measured quantities can then be compared to analogues of Sgr in simulations to test for consistency of the $\Lambda$CDM model. Improved uncertainties can also be used to better constrain a dark matter annihilation or decay signal \cite{2023MNRAS.524.4574E}.

We reiterate that in our analysis, we assume that the core of Sgr is in dynamical equilibrium, and that the presence of tidal disruption in the region of our studied sample may be neglected. Of course, we do not have definitive proof that this is the case. However, there is good motivation that our assumptions are at least a good approximation, given the following observations/theoretical results. First, in the sample that we identify, we do not detect statistically significant presence of streaming motion in the velocities in any of the coordinate directions we consider. Second, the observed velocity distribution is nearly gaussian in each of the coordinate directions, indicative of a system in near equilibrium. Third, from a theoretical perspective, the true dark matter mass for Sgr analogues in simulations has been shown to be well-determined assuming that the core of the system is in equilibrium and may be modeled using the jeans equations \citep{2022ApJ...941..108W,2025arXiv250418617T}. For these reasons, we believe that equilibrium is a good approximation to the system for the sample that we consider. We also make the assumption of sphericity, which is known to not be the case for Sagittarius. Previous studies have found $\sim 30$\% differences between spherical and non-spherical jeans models \citep{2015MNRAS.446.3002B, 2018MNRAS.474.1398G, 2024ApJ...970....1V}.

\par This paper provides a template for future studies of dSphs with full 6D position and velocity information. This will be possible in the future with precision measurements of distances to RR Lyrae in the Sgr core. 
Previous studies along these lines have shown that including three-dimensional positions with line-of-sight velocities of stars are expected to improve measurements of the velocity anisotropy parameter~\citep{2014MNRAS.440.1680R}. Full six-dimensional phase space coverage of stars in dSphs will allow for an unprecedented analysis of the dynamical state of the dark and luminous
mass in dwarf galaxies.

\section{Acknowledegments} 
The authors would like to thank Matthew Walker for his coding contributions, as well as Oz and Mac for their support. LS acknowledges support from DOE Grant desc0010813. This work has made use of data from the European Space Agency (ESA) mission {\it Gaia} (\url{https://www.cosmos.esa.int/gaia}), processed by the {\it Gaia} Data Processing and Analysis Consortium (DPAC, \url{https://www.cosmos.esa.int/web/gaia/dpac/consortium}). Funding for the DPAC has been provided by national institutions, in particular the institutions participating in the {\it Gaia} Multilateral Agreement.
Portions of this research were conducted with the advanced computing resources provided by Texas A\&M High Performance Research Computing.

\section{Data Availability} 
The data used in this article are publicly available from the European Space Agency (ESA) mission \textit{Gaia} and can be accessed through the \href{https://gea.esac.esa.int/archive/}{Gaia Archive}. The resulting table of Sagittarius candidates used in this analysis is included as additional material in the file Sgr\_candidates.txt.


\bibliographystyle{mnras}
\bibliography{Sgr6D}



\appendix
\section{Gaia Query}\label{app:queries}

\noindent 
Query:
\newline
\texttt{
SELECT source\_id, ra, ra\_error, dec, dec\_error, pmra, pmra\_error, pmdec, pmdec\_error, parallax, parallax\_over\_error, phot\_g\_mean\_mag, phot\_rp\_mean\_mag, phot\_bp\_mean\_mag, radial\_velocity, radial\_velocity\_error \\
FROM gaiadr3.gaia\_source AS gaia \\
   WHERE parallax < 1 \\
   AND gaia.ra > 277 AND gaia.ra < 290  \\
   AND gaia.dec < -24 AND gaia.dec > -37
}

\section{Additional Analyses}\label{app:extraanalysis}

\begin{figure}
    \centering
    \includegraphics[width=0.7\textwidth]{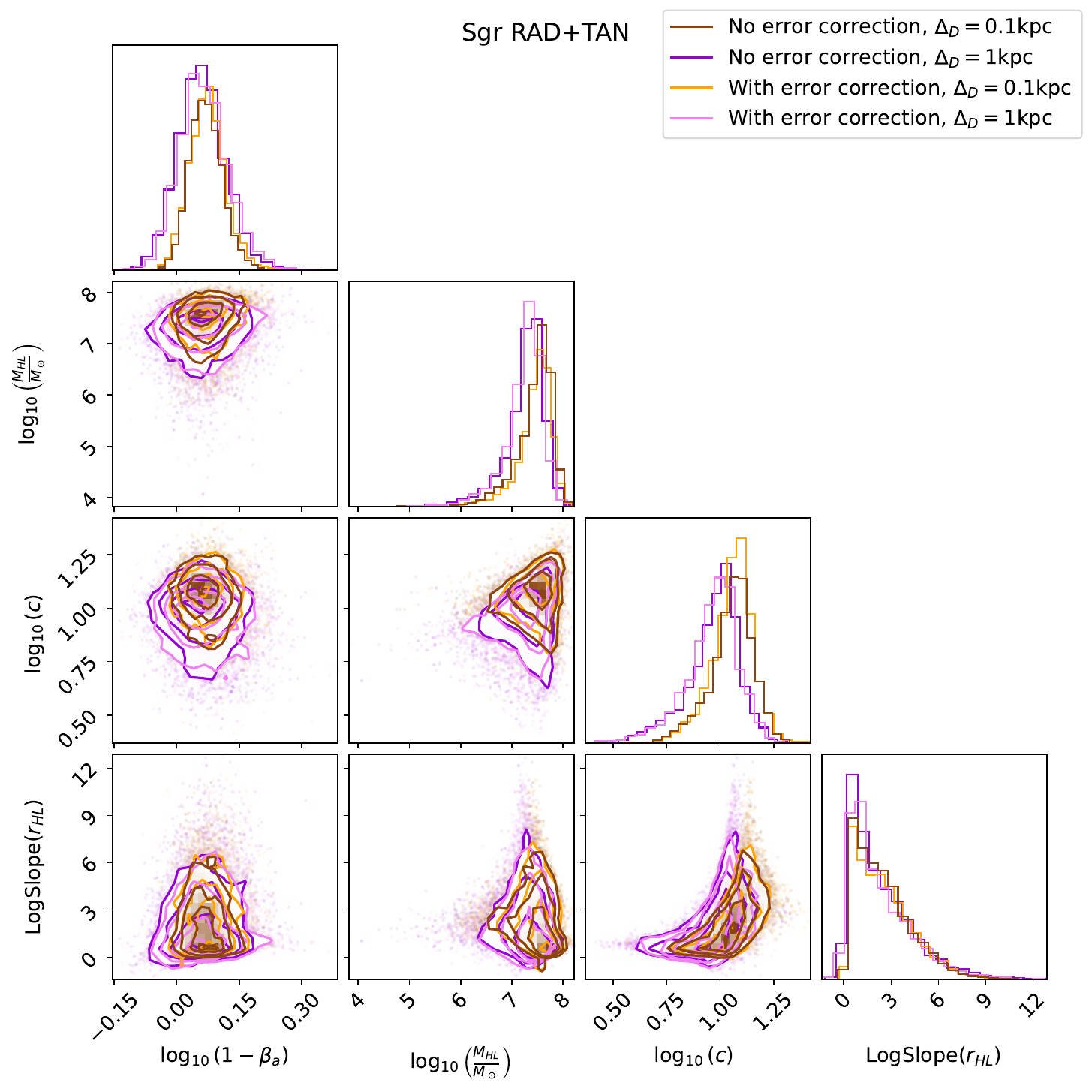}
    \caption{Posteriors for a combined radial and tangential Jeans analysis. Brown and dark purple contours correspond to an analysis with no systematic error corrections and $\Delta_D=0.1$kpc, $\Delta_D=1$kpc respectively. Orange and pink contours correspond to an analysis with the systematic error correction for $\Delta_D=0.1$kpc, $\Delta_D=1$kpc respectively.
    The parameters are velocity anisotropy in the form $\log_{10}(1-\beta_a)$, mass contained within the half light radius $\log_{10}(M_{HL}/M_\odot)$, concentration $\log_{10}(c)$, and log slope at the half light radius LogSlope($r_{HL}$).}
    \label{fig:syserrcorrect}
\end{figure}

Gaia proper motion errors are known to be underestimated, since they do not include a component for systematics \citep{2021A&A...649A...2L}. \cite{2021MNRAS.505.5978V} quantify this effect with a correction factor added in quadrature. Taking a systematics correction for proper motion as suggested in \cite{2021MNRAS.505.5978V}, we perform a RAD+TAN Jeans analysis with both $\Delta_D=1$ kpc, $\Delta_D=0.1$kpc. 
Figure \ref{fig:syserrcorrect} compares the resulting posteriors to their counterpart without a systematics correction. No significant change is observed when the systematic error correction is taken into account. 

\bsp	
\label{lastpage}
\end{document}